\documentclass{Rinton-P9x6}
\def\transp#1{{#1}^\tau}

\begin{document}

\title{LOCC protocols for entanglement transformations}

\author{G. Mauro D'Ariano$^{a,b}$ and  Massimiliano F. Sacchi$^{a}$}

\address{(A) Dipartimento di Fisica `A. Volta', Universit\`a di Pavia 
and Unit\`a INFM\\ via A. Bassi 6, I-27100 Pavia, Italy} 
\address{(B) Department of Electrical and Computer Engineering\\
Northwestern University, Evanston, IL 60208, USA}

\maketitle

\abstracts{We construct the protocols to achieve probabilistic
and deterministic entanglement transformations for bipartite pure
states by means of local operations and classical communication. A new
condition on pure contraction transformations is provided.}

The transformation of entangled states by means of local operations
and classical communication (LOCC) is a key issue in quantum
information processing. In fact, increasing entanglement by means of
LOCC with some probability is crucial in practice, since losses and
decoherence have detrimental effects in the establishment of
entanglement at distance.

In this paper we give a short and simple proof of
Lo-Popescu theorem\cite{Lo-Popescu}. Then, we provide a new necessary
condition for pure-contraction transformations. Finally, 
we construct explicitly the LOCC protocols to achieve 
deterministic and probabilistic transformations for
bipartite pure states.

We introduce now the main notation.  Given a linear
operator $O$ we denote by $O^\dag $, $O^*$, $\transp{O}$, and $O^\ddag
$ the hermitian conjugate, the complex conjugate, the transpose, and the
Moore-Penrose inverse of $O$, respectively.   Recall that $O^\ddag$ is the
unique matrix that satisfies $OO^\ddag O=O$, $O^\ddag OO^\ddag
=O^\ddag $, $OO^\ddag $ and $ O^\ddag O $ hermitian. Notice also that
$OO^\ddag\equiv P_O$ is the orthogonal projector over $\hbox{Rng}(O)$,
whereas $O^\ddag O\equiv P_{O^\dag}$ is the orthogonal projector over
$\hbox{Rng}(O^\dag)\equiv \hbox{Supp}(O)$.  We write the
singular value decomposition (SVD) of $O$ as $O=X_O\Sigma _O Y_O $,
where $\Sigma _O$ denotes the diagonal matrix whose entries are the
singular values $\sigma _i (O)$ of $O$ taken in decreasing order, and
$X_O,\ Y_{O}$ are unitary.  We write
\begin{eqnarray}
|A\rangle\!\rangle \equiv \sum_{i,j}a_{ij} |i \rangle _1\otimes |j \rangle _2   
\;\label{kket}
\end{eqnarray}
for the bipartite pure states on the Hilbert space
${\cal H}_{1}\otimes {\cal H}_{2}$, 
where $\{|i \rangle _1\}$ and $\{|j \rangle _2\}$ are two 
orthonormal bases for ${\cal H}_{1}$ and 
${\cal H}_{2}$. One can easily check the relation $
A\otimes B |C\rangle\!\rangle =|AC\transp{B}\rangle\!\rangle $. 
Finally, we use the notation  $A\prec B$ for Hermitian 
operators $A$ and $B$ to denote the majorization relation\cite{bhatia} 
$\hbox{eigv}(A)\prec\hbox{eigv}(B)$, and 
in the same fashion we will write $A\prec^w B$ and $A\prec_w
B$ for super- and sub-majorization\cite{bhatia}.

\par We give now a simple proof of Lo-Popescu theorem\cite{Lo-Popescu}. 
\par\noindent 
{\bf Theorem 1.} All LOCC on a pure bipartite entangled state
$|\Psi\rangle\!\rangle $ can be reduced to a contraction by Alice and 
a unitary transformation by Bob. This is due to the equivalence
of any Bob contraction $M$ with the Alice contraction $N$
assisted by Bob's unitary transformation $U$ as follows
\begin{eqnarray}
I\otimes M|\Psi\rangle\!\rangle =N\otimes U|\Psi\rangle\!\rangle \;,\label{LoP}
\end{eqnarray}
where
\begin{eqnarray}
N =K_{M\transp{\Psi }} M 
K_{\Psi } \;,\;\qquad 
U =K^\dag _{M\transp{\Psi }}K^\dag _{\Psi }\;,\label{due}
\end{eqnarray}
and $K_O$ is the unitary operator achieving the transposition of the
operator $O$, namely 
$\transp{O} = K_O O K_O^* $. 
\par\noindent 
{\bf Proof.} To prove that every LOCC can be reduced to an Alice contraction
and a Bob unitary transformation it is sufficient to prove
 equivalence (\ref{LoP}), since: {\em a)}  all possible elementary
LOCC in a sequence will be reduced to an Alice contraction
and a Bob unitary; {\em b)} the product of two contractions is a
contraction; {\em c)} unitary transformations are particular cases of
contraction. 
\par\noindent  Given the SVD of any linear
operator $O$ one has 
\begin{eqnarray}
\transp{O} = Y_\transp{O} \Sigma_O X_\transp{O}= (Y_\transp{O}
X_O^\dag)O 
(\transp{Y_O} X_O^\dag )^* \equiv   K_O O K_O^*  
\;,\label{ko}
\end{eqnarray}
with $K_O=Y_\transp{O} X_O^\dag$. Hence 
\begin{eqnarray}
\Psi \transp{M} = \transp{(M\transp{\Psi })}= K_{M\transp{\Psi }}
(M\transp{\Psi }) K^*_{M\transp{\Psi }} =K_{M\transp{\Psi }}M 
K_{\Psi }\Psi K _\Psi ^* K^*_{M\transp{\Psi }}
\;. 
\end{eqnarray}
Then one gets Eq. (\ref{LoP}) with  $N $ and $U$ given as 
in Eq. (\ref{due}).  
\par The main theorem on entanglement transformations is the following.
\par\noindent{\bf Theorem 3.}
The state transformation $|A\rangle\!\rangle \to|B\rangle\!\rangle $ is possible by LOCC iff 
\begin{equation}
AA^\dag\prec^w pBB^\dag \;,\label{MajBA}
\end{equation}
where $p\le 1$ is the probability of achieving the
transformation. A necessary condition to be satisfied is
$\hbox{rnk}(A)\ge\hbox{rnk}(B)$. 
In particular, the transformation is deterministic
($p=1$) iff $AA^\dag\prec BB^\dag$.  

\par\noindent Theorem 3 unifies the results of Nielsen\cite{nielsen99} and 
Vidal\cite{vidal99}. 
\par In the following we provide a new necessary condition for the case of
pure-contraction transformation, namely we prove:
\par\noindent{\bf Theorem 3.} 
If there is a {\em pure} LOCC 
that achieves the state transformation $|A\rangle\!\rangle \to|B\rangle\!\rangle $ 
with probability $p$, we must have
\begin{equation}
pBB^\dag\prec_w AA^\dag.\label{MajBA2}
\end{equation}
\par\noindent{\bf Proof.} According to theorem 1, the pure LOCC 
transformation $|A\rangle\!\rangle  \to  |B\rangle\!\rangle $ occurring with probability $p$
is given by 
\begin{equation}
M\otimes U|A\rangle\!\rangle =\sqrt{p}|B\rangle\!\rangle \;,\label{LOCC}
\end{equation}
and we need to have $
MA\transp{U}=\sqrt{p}B$. Using the SVD of $A$ and $B$ one has $
\tilde{M}\Sigma_A\tilde{U}=\sqrt{p}\,\Sigma_B $, 
with $\tilde{M}=X_B^\dag MX_A $ and 
$\tilde{U}=Y_A\transp{U}Y_B^\dag $. Then 
$\tilde{M}\Sigma_A^2\tilde{M}^\dag=p\Sigma_B^2$, namely
\begin{equation}
\sum_k S_{kl}\sigma_k^2(A)=p\sigma_l^2(B)\;,\label{S}
\end{equation}
where $S_{kl} \doteq |\langle  l|\tilde{M}|k \rangle |^2$ is a sub-stochastic
matrix, since 
\begin{eqnarray}
\sum_k S_{kl}=&&\langle  l|\tilde{M}\tilde{M}^\dag|l \rangle \le
||M^\dag||^2\le
1\;,
\nonumber \\ 
\sum_l S_{kl}=&&\langle  k|\tilde{M}^\dag\tilde{M}| k \rangle 
\le ||M||^2\le  1\;.
\label{}
\end{eqnarray}
This proves that Eq. (\ref{MajBA2}) is a necessary condition 
for transformation (\ref{LOCC}).

\par To construct the explicit protocols that realize entanglement
transformations we 
will use the following lemma:
\par\noindent{\bf Lemma 1.}
$x\prec^w y \iff $ for some $v\quad x\prec v $ and $v\geq y$, 
\par\noindent along with Uhlmann theorem:
\par\noindent{\bf Theorem 2.}
For Hermitian operators $C$ and $D$ one has
$C\prec D$ if and only if there is a probability distribution
$p_\lambda $ 
and unitaries $W_\lambda $ such that 
\begin{equation}
C=\sum_\lambda  p_\lambda  W_\lambda  D W_\lambda ^\dag\;.
\label{ranW}\end{equation} 
\par\noindent For Lemma 1 a probabilistic transformation can always be
performed through two steps:
a deterministic transformation $|A\rangle\!\rangle  \to |Q\rangle\!\rangle $, followed by a
pure-contraction $|Q\rangle\!\rangle  \to |B\rangle\!\rangle $ that occurs with probability $p$. 
\par For the deterministic transformation $|A \rangle\!\rangle  \to |Q\rangle\!\rangle $, 
one needs to find the contractions $M_\lambda $ and
the unitaries $U_ \lambda $ versus the operators
$W_\lambda $ of Eq. (\ref{ranW}), where $C=AA^\dag $ and $D=QQ^\dag $, 
such that 
\begin{eqnarray}
M_\lambda\otimes U_\lambda |A\rangle\!\rangle =\sqrt{q_\lambda} |Q\rangle\!\rangle \;.
\label{mza}
\end{eqnarray}
The general solution of Eq. (\ref{mza}) 
is given by
\begin{eqnarray}
M _\lambda =\sqrt{q_\lambda } QU_\lambda ^* A^\ddag +
N_\lambda (1- AA^\ddag )   
\;.\label{ml}
\end{eqnarray}
To guarantee that $M_\lambda $ is a contraction 
we can always take 
\begin{eqnarray}
U_\lambda ^* =Y_Q^\dag X_Q^\dag W_\lambda X_A Y_A
\;,\qquad N_\lambda =0\;.\label{zl}
\end{eqnarray}
In fact from Eqs. (\ref{ml}), and (\ref{zl}) and using 
Eq. (\ref{ranW}) one has
\begin{eqnarray}
\sum _\lambda {M^\dag _\lambda M_\lambda }&=&
\sum_\lambda q_\lambda (A^\ddag )^\dag Y_A^\dag X_A ^\dag W^\dag _\lambda 
X_Q Y_Q  Q^\dag 
Q  Y_Q^\dag X_Q^\dag W_\lambda X_A Y_A A^\ddag  \nonumber \\&= & 
\sum_\lambda q_\lambda (A^\ddag )^\dag Y_A^\dag X_A ^\dag W^\dag _\lambda 
Q Q^\dag  W_\lambda X_A Y_A A^\ddag   
\nonumber \\&= & 
(A^\ddag )^\dag Y_A^\dag X_A ^\dag A A^\dag X_A Y_A A^\ddag
  \nonumber \\&= & 
(A^\ddag )^\dag  A^\dag A A^\ddag   =
(A A^\ddag )^\dag  A A^\ddag   = A A^\ddag   = P_A 
\;. 
\end{eqnarray}
The completeness of the measurement can be guaranteed by the further
contraction $ M_0=V (I -A A^\ddag )$ 
where $V$ is an arbitrary unitary operator.
\par Hence, 
given explicitly Eq. (\ref{ranW}) 
 one can perform the contractions $M_\lambda $
and the unitaries $U_\lambda $ to achieve the entanglement
transformation. The problem of looking for a POVM with minimum number
of outcomes (thus minimizing the amount of classical information sent
 to Bob's side) is reduced to find the transformation (\ref{ranW}) with
minimum number of unitaries. 
One can resort to a constructive algorithm to
find a bistochastic matrix $D$ 
which relates the vectors $\vec \sigma ^2 _A$ and $\vec \sigma ^2 _Q$ 
of the singular values of $A$ and $Q$, namely 
$\vec \sigma ^2 _A= D \vec \sigma ^2 _Q $. 
Then Birkhoff theorem allows to write $D$ as a convex
combination of permutation matrices 
$D=\sum _\lambda q_\lambda \Pi _\lambda $.  
In terms of $\Sigma _A$ and $\Sigma_Q$ one 
has 
\begin{eqnarray}
\Sigma_A^2 =\sum_\lambda q_\lambda \Pi ^\dag _\lambda 
\Sigma_Q^2 \Pi  _\lambda 
\;\label{pil}
\end{eqnarray}
where $\Pi _\lambda =\sum _l |l \rangle \langle \Pi _\lambda (l)|$.
In this way one obtains Eq. (\ref{ranW}), with $W_\lambda =X_Q \Pi
_\lambda X_A^\dag $. Using Eqs. (\ref{ml}) and (\ref{zl}) for 
$M_\lambda $ and $U_ \lambda $ one recovers the
result of Ref. 5.
Notice that Caratheodory's theorem always allows to reduce the number of
permutations in Eq. (\ref{pil}) to $(d-1)^2+1$, for $d$-dimensional
Alice's Hilbert space.    

\par The second part of the protocol, namely the contraction which
provides the state $|B\rangle\!\rangle $ from $|Q \rangle\!\rangle $, is present only for
probabilistic transformations.  It is a pure contraction given by
\begin{eqnarray}
N \otimes V |Q\rangle\!\rangle =|N Q \transp{V} \rangle\!\rangle =
\sqrt p |B\rangle\!\rangle \;, 
\end{eqnarray}
where  
$N=\sqrt p X_B \Sigma _B \Sigma _Q^\ddag X_Q^\dag $
and 
$\transp{V}=Y_Q^\dag Y_B $. In fact 
\begin{eqnarray}
N Q \transp{V}=\sqrt p X_B \Sigma _B \Sigma _Q ^\ddag \Sigma _Q Y_B
= \sqrt p X_B \Sigma _B  Y_B = \sqrt p B
\;,
\end{eqnarray}
where we used the fact that $\Sigma _B \Sigma _Q^\ddag \Sigma _Q =
\Sigma _B $, since for lemma 1 one has $\Sigma _Q^2 \geq p \Sigma _B^2$. 

\section*{Acknowledgments}This work has been jointly founded by the EC under the program 
ATESIT (Contract No. IST-2000-29681) and by the USA Army Research
Office under MURI Grant No. DAAD19-00-1-0177.

\end{document}